\documentclass[11pt]{article}
 
\usepackage{epsfig}
\usepackage{amsmath}
\usepackage{amssymb}
 
\begin{document}


\renewcommand{\thefootnote}{\fnsymbol{footnote}}

\title{ A semi-variational approach to QCD at finite temperature and baryon density} 

\author{ Fabrizio Palumbo\\
\small\it  INFN, Laboratori Nazioni Frascati, \\ 
\small\it  P.~O.~Box 13, I-00044 Frascati, Italy \\ 
\small\tt Fabrizio.Palumbo@lnf.infn.it    
}

\date{\today}
\maketitle

\begin{abstract}
\noindent

Recently a new bosonization method has been used to derive, at zero fermion density, an effective action for relativistic
field theories whose partition function is dominated by fermionic composites,  chiral mesons in the case of QCD. 
This approach shares two important features with variational methods: the restriction to the subspace of the 
 composites, and the determination of their structure functions 
  by a variational calculation. But unlike standard variational methods it treats  excited
  states on the same footing as the ground state.
  
 I extend this method  including states of nonvanishing 
fermion (baryon) number and derive an effective action for QCD at finite temperature and
baryon density. I test the result on a four-fermion interaction model.
\end{abstract}

\vfill\eject

\renewcommand{\thefootnote}{\arabic{footnote}}
\setcounter{footnote}{0}

\section{Introduction}

Increasing temperature and baryon density hadronic matter  is expected to undergo one or more
crossovers and/or
phase transitions. Increasing temperature at zero baryon number one might/should meet a phase in which quarks
cohexist with hadronic (possibly colored) states~\cite{Lerr}. Increasing baryon density at fixed temperature
one should meet a similar phase and  possibly  a
color superconducting phase~\cite{Colo} due to a weak attractive channel between quarks of different
colors. These new states of matter should be at least partially accessible to  experimental 
investigation in  heavy ion collision experiments.

Understanding the behavior of hadronic matter at high temperature and baryon density is relevant  for the study of early 
Universe and  neutron stars. But its theoretical properties  can be studied only nonperturbatively and 
the lattice approach, the most powerful tool for first principles, nonperturbative studies, 
is affected in the case of finite density QCD by the well known sign problem.

Some progress was achieved recently~\cite{Forc, Elia}  by simulations at imaginary 
chemical potential. Other interesting results were obtained~\cite{Fodo} within a modified version  of 
the Glasgow re-weighting technique  and by an  approach which makes use of a Taylor expansion in the chemical
potential in the small $\mu/T$
region~\cite{Allt}.

 At last a new approach to simulate QCD at finite temperature and baryon density was
 developed~\cite{Azco}, which  resembles in some aspects that of the  imaginary chemical potential, 
 but seems to have a wider  range of applicability~\cite{Azco1}. 
 
 The magnitude of quark masses has large effects in numerical simulations.
 Several arguments lead to the expectation  
 that the evaluation of the fermion determinant is more stable the larger fermion masses 
are~\cite{Aloi}. These arguments are relevant to  the present work,
as explained at the end of Section VI.

To tackle the problem of QCD at finite temperature and baryon density I extend  a new method 
constructed to treat fermionic systems
whose partition function is dominated by fermionic composites. This is certainly the case of QCD
at low temperature and baryon density, in which the  relevant degrees of freedom are mesons and nucleons, 
and also at high temperature and baryon density according to the expectations reported above.
  This method was first 
 developed in the framework of many-body nonrelativistic theories ~\cite{Palu} and then applied to relativistic field 
 theories~\cite{Cara} at finite temperature and zero fermion density. In the case of QCD, neglecting nucleons it amounts
  to a bosonization.
 The heuristic motivation is that reformulation of a theory in terms of fields related to physical degrees of freedom should
make it simpler. 
The starting point of this approach  is the partition function in operator
form, namely the trace of the transfer matrix in the Fock space of the 
fermions. The physical assumption
of composite dominance is then implemented by restricting the trace to fermion
composites. This requires an
approximation of  a projection operator on  the subspace of the composites, 
the approximation being the better,
the higher the number of fermion states (called index of nilpotency) in the composites. The approximate 
projection operator is constructed
in terms of coherent states of composites, and evaluation of the trace, which is done exactly, 
generates  a bosonic action in terms
of the holomorphic variables appearing in the coherent states.

The structure functions of the composites are determined by a variational procedure. So this approach
shares two important features with variational methods: The restriction to a subspace of the Fock space of fermions, 
  the space of  chiral mesons in the case of QCD, and the determination of their structure  functions 
  by a variational calculation. But unlike standard variational methods in the present one excited
  states are treated  on the same footing as the ground state.

 The  utility of variational methods and bosonization has been widely appreciated in the theory of many-body systems.
  But their potentiality has also been considered in the framework of relativistic field theories, in particular gauge theories, 
  for example by R.~Feynman~\cite{Feyn} who, however, was skeptical about their practical applicability, and recently
   in connection with QCD at high baryon density~\cite{Wilc}.

The approach just outlined is compatible with any regularization. But
 in  gauge theories the effective action  of the composites will
involve vacuum expectation values of  invariant functions of gauge fields
which cannot be evaluated within the present framework. Therefore a lattice formulation 
was adopted in order to be able to  to extract such expectation values from numerical simulations.
One is then confronted with  the well 
known difficulty with chiral invariance, which can only in part be overcome by 
using Kogut-Susskind fermions. However the method can, at least in principle,
be used with any other lattice regularization~\cite{Frez} for which a transfer matrix has been explicitly
constructed.

 The formalism of the transfer matrix 
does not treat time and space in a symmetric way, and therefore Euclidean
invariance of the bosonic action must  be checked a posteriori. All other
symmetries are instead respected.

Before outlining the extension of this approach I will resume what has been already done.
The validity of the method was tested on  a model with a 4-fermion interaction in 3+1
dimensions:  Euclidean invariance was recovered in the continuum limit and all the known results 
in the boson sector were exactly 
reproduced, namely condensation of a 
composite boson with the right mass, which breaks the discrete chiral
invariance of the model.
In addition  the structure function of the composite was determined, and its  radial factor, in a polar representation,
turned out to be identical with that
of the Cooper pairs of the BCS model of superconductivity. 

To study QCD at finite temperature and  baryon density as a first step
I must introduce
quark states in the presence of mesons, namely I must construct {\it a Fock space containing composites and
their constituents  avoiding  double counting}. A
similar but more difficult problem has been considered since a long time: given a Lagrangian
which generates bound states, how to replace it by a physically equivalent Lagrangian in which
bound states and constituents are treated on equal footing~\cite{Houa}. I solve my problem defining
quasiquark states in such a way  that quasiquark-quasiantiquark states  are orthogonal to meson states.
This constraint corresponds to the condition on the wave function renormalization of  composite
particles in the Lehman spectral representation of composite operators~\cite{Houa}. 

The next step, the  explicit introduction of baryons and antibaryons constructed in terms of
quasiquarks and quasiantiquarks is desirable
but not necessary in a variational calculation, because a space of mesons, quasiquarks and quasiantiquarks obviously contains 
 baryons and antibaryons. 
Therefore in the present paper I will not explicitly include in the partition function  
baryonic states. For a further simplification, which will be removed in work in progress, 
 I will exclude antiquarks, so that my variational space contains mesons and baryons. 
 This amounts to neglect virtual baryons-antibaryons, and it is justified for not too high temperature and baryon 
 density. 
{\it In the resulting effective action the expectation value of the chiral sigma field provides a mass to the quasiquarks.
On the ground of the arguments concerning the effects of quark masses on numerical simulations
quoted above~\cite{Aloi}, I hope  that  numerical simulations with such effective action will be more stable}. 
I will also investigate  in a separate work  the  possiblilty of analytical expansions.

I again test the method on a four-fermion interaction model, reproducing the known results in the fermion sector,
namely existence of a free fermion whose mass is half that of the composite boson, and chiral symmetry restoration
with increasing fermion density. The mechanism of this restoration is that quasifermions occupy the
lowest energy states, from zero energy up to a maximum energy increasing with density, progressively
depleting the condensate.

The paper is organized in the following way. In Section 2 I report the general formalism at zero baryon density,
in section 3 I define quasiquark states and the approximate projection operator in 
the subspace
of mesons and quasiquarks, in section 4 I derive a first form of their effective action. In Section
5  I apply this action to the study of the four-fermion interaction model, deriving the results described above.
In section 6 I derive a second form of the effective action, which has a more transparent interpretation, and
allows a crosscheck of the accuracy of the approximation for the projection operator by comparison of the
results for the four-fermion interaction model, which coincide with those obtained by the first effective action.
In Section 7 I summarize my results with an outlook to possible applications.

\section{General formalism at zero baryon density}

To make the paper reasonably selfcontained  I report the general formalism  developed for 
relativistic field theories of fermions  in the presence of composites dominance at zero fermionic 
density~\cite{Cara}. I make
 only a small modification: I do not fix the gauge, because this would  prevent  numerical simulations
 and complicate the treatment due to the need to enforce  the Gauss constraint in Fock space. 
 
 The
 starting point of this formalism is  the standard expression of the  partition function of QCD
 in terms of the transfer matrix
\begin{equation}
\mathcal{Z} = \int [dU] \exp \left[- S_G(U) \right]\mbox{Tr}^{\mbox{F}} 
\left\{ \prod_{t=0}^{L_0-1}
 \left( \hat{T}^\dagger_t  \, {\hat V}_t \exp(\mu \, \hat{n}_B) \hat{T}_{t+1}\right) \right\}\, \label{part}
\end{equation}
where $L_0$ is the number of links in the temporal direction, $S_G$ is the gluon action and
\begin{eqnarray}
\hat{T}_t  &=&\exp [ -\hat{u}^{\dagger} M_t\, \hat{u} - \hat{v}^{\dagger}  
M^T_t \hat{v} ] 
\exp[\hat{v} N_t  \, \hat{u}] 
\nonumber\\
{\hat V}_t &=& \exp  [ \hat{u}^{\dagger}\ln  U_{0,_t}\, \hat{u} + \hat{v}^{\dagger}  
U_{0,t}^* \, \hat{v} ]  \,.
\end{eqnarray}
The  $ U_{\mu,t}$ are matrices whose matrix elements are the link variables at Euclidean 
time "t"
\begin{equation}
(  U_{\mu,t} )_{{ \bf x}_1,{ \bf x}_2 }
= \delta_{{ \bf x}_1,{ \bf x}_2 } U_{\mu,t} ( {\bf x}_1)\,.
\end{equation}
 Because the formalism treats asymmetrically time
 and space, I use ${\bf boldface}$ letters,
as $\mathbf{x}$, to denote spatial coordinates, and {\it italic} 
letters to denote space-time coordinates: ${\it x} =(t,\mathbf{x})$.
$\hat{u}^\dagger_i$ and $\hat{v}^\dagger_i$ are, respectively, creation 
operators of quarks and antiquarks in state $i$, obeying canonical anti-commutation relations.
$\mbox{Tr}^F$ is the trace  over the Fock space of quarks, $\mu$ is the chemical potential and 
$\hat{n}_B$ the baryon 
number operator. 
The matrices $M_{t}$ ($M_t^T$ being the transposed of $M_t$) and $N_t$ are functions of the  spatial 
link variables at time $t$. 
They depend on the regularization adopted for the fermions, but what follows 
is not affected by their explicit expressions, which are  reported 
in  Appendix~\ref{B} for  Wilson and Kogut-Susskind fermions in the flavor basis.

I include in the gluon action the term
\begin{equation}
\delta S_G =  \sum_t - 4 \, \mbox{tr}_{-} M_t 
\end{equation}
which  comes from transformations on the fermion fields going from the functional form to this operator form 
of the transfer matrix. 
I introduced the notation, which I will use  for any matrix $\Lambda$
\begin{equation}
\mbox{tr}_{\pm} \Lambda= \mbox{tr}\left( P_0^{(\pm)} \Lambda \right) \,.
\end{equation}
The  operators $P_0^{(\pm)}$, which  project on the quark antiquark components of the quark field
are defined in  Appendix~\ref{B}.  $\mbox{tr}_{\pm}  $ is the trace over quarks or
antiquarks intrinsic quantum numbers and spatial coordinates (but not over time). 

The expression~(\ref{part}) for the partition function was given by Lus\"cher ~\cite{Lusc} in the gauge $U_0 \sim 1$,
in which $V_t =1$ (but one has to impose the Gauss constraint in the Fock space of fermions).

Under the assumption that at low energy the partition function is dominated by chiral mesons,
  the trace in the Fock space can be restricted to them. The restricted partition function can be written
introducing an operator  $\mathcal{P}_{\mbox{m}}$ which projects in the subspace of mesons
\begin{equation}
\mathcal{Z}_{\mbox{mesons}} = \int dU \exp \left[- S_G(U) \right]
\mbox{Tr}^{\mbox{F}} \left\{ \prod_{t=0}^{N_0-1}
 \left( \mathcal{P}_{\mbox{m}} \, \hat{T}^\dagger_t  \, {\hat V}_t \exp(\mu \, 
 \hat{n}_B) \hat{T}_{t+1}\right) \right\}\,.
\end{equation}
To construct this projector,  meson creation operators are introduced
\begin{eqnarray}
\hat{\Phi}_{\mathbf{x},K}^\dagger=\hat{u}^\dagger\Phi_{\mathbf{x}K}^\dagger
\hat{v}^\dagger=
\sum_{ij}\hat{u}^\dagger_i(\Phi_{\mathbf{x}K}^\dagger)_{ij}\hat{v}^\dagger_j\, ,
\end{eqnarray}
where $\mathbf{x}$ represents their spatial coordinate, 
 $K$ their  quantum numbers, like radial excitations, spin, flavor, etc, and
 $\Phi_{\mathbf{x}K}$ their structure functions ({\it wave functions}).  

Since  fermion creation operators are nilpotent, composite creation operators $\hat{\Phi}^{\dagger}_{{\bf x},K}$
can be classified according  to their index of nilpotency, which is the highest integer exponent $\Omega$ such that 
\begin{equation}
\left (\hat{\Phi}^\dagger\right)^\Omega \neq 0 \,.
\end{equation}
$\Omega$ counts the number of fermion states in the composite.
By analogy with  systems of elementary bosons coherent states of mesons can be constructed
\begin{equation}
|\phi\rangle =
\exp\left(\sum_{\mathbf{x},K}\phi_{\mathbf{x}K}
\hat{\Phi}^\dagger_{\mathbf{x}K}\right)\,|0\rangle\, ,
\end{equation}
where the $\phi_{\bf{x}K}$'s are  holomorphic variables. But  since  composites operators do not obey canonical 
commutation relations, the properties of their coherent states can 
  differ from those of canonical bosonic coherent states. For instance the basic property of
coherent states cannot be exactly satisfied
\begin{equation}
\hat{\Phi}_{\bf{x} K} | \phi \rangle \neq \phi_{\bf{x} K} | \phi \rangle  \label{basic}\,.
\end{equation}
 However, if the index of nilpotency of the composites
is large enough, the composites system resembles a canonical bosonic system,
and  the properties of canonical boson coherent states 
will approximately hold for the composite coherent states, as shown in detail in Refs. \cite{Palu, Cara}.

 Hence, under the assumption that the composite operators 
which dominate the
partition function have a large index of nilpotency, an approximate projection operator in the Fock space of the fermions
can be defined
\begin{equation}
\mathcal{P}_{\mbox{m}} = \int  \left[ { d\phi d\phi^* \over 2 \pi i} \right]\,
\langle\phi|\phi\rangle^{-1} \, |\phi\rangle \langle\phi| \, ,
\label{projector}
\end{equation}
where 
\begin{equation}
\left[ { d\phi d\phi^* \over 2 \pi i} \right]  =\prod_{\mathbf{x},K} 
\left[ { d\phi_{\mathbf{x}K} d\phi^*_{\mathbf{x}K} \over 2\pi i } \right].
\end{equation}
It is important to observe  that the space selected by this operator  includes 2 physically equivalent
states obtained for $\phi=0, \infty$. They correspond to a completely empty or filled lattice.

The scalar product of coherent states appearing in the definition of the projection operator is
\begin{equation}
\langle\phi|\phi'\rangle =
{\det}_{+} \left[\, \mathcal{I}\:+\: (\phi\cdot\Phi^\dagger)\,
(\phi'^*\cdot\Phi) \right]\, ,
\end{equation}
where
\begin{equation}
 \phi\cdot\Phi^\dagger =\sum_{\mathbf{x},K}\phi_{\mathbf{x}K}
 \Phi_{\mathbf{x}K}^\dagger
 \end{equation}
 and for any matrix $\Lambda$
\begin{equation}
 { \det}_{\pm} \Lambda = \det(P_0^{\pm}\Lambda) \,.
\end{equation}
 $\mathcal{I}$ is the identity in the space of all the matrices. I remind that the entries of
these matrices do not include time. By a little abuse of notation I will often write "1" instead of $\mathcal{I}$.

With periodic boundary conditions for the gauge fields, the partition 
function is
\begin{equation}
\mathcal{Z} = \mathrm{Tr}^F\,\left\{ \hat{T}^\dagger_0\,{\hat V}_0 \,\exp(\mu \,n_B)  
\hat{T}_{1}^{\vphantom{\dagger}}\,
 \hat{T}^\dagger_1 \, {\hat V}_1 \,\,\exp(\mu \,n_B)  \cdots \hat{T}^\dagger_{L_0-1}\,
{\hat  V}_{L_o-1}\exp(\mu \,n_B) \hat{T}_{0}^
{\vphantom{\dagger}} \right\}
\end{equation}
while its restriction to mesons is
\begin{eqnarray}
\mathcal{Z}_{\mbox{mesons}} &=& \mathrm{Tr}^F\,\left\{ {\cal P}_{\mbox{m}}\,
\hat{T}^\dagger_0\, {\hat V}_0 \,
\hat{T}_{1}^{\vphantom{\dagger}} \, {\cal P}_{\mbox{m}}\,\hat{T}^\dagger_1 \,{\hat V}_1
\cdots {\cal P}_{\mbox{m}}\,
\hat{T}^\dagger_{L_0-1}\,{\hat V}_{L_0 -1}\hat{T}_{0}^{\vphantom{\dagger}} \right\}
\nonumber\\
&=& \int \prod_{t=0}^{L_0-1} \left[ \frac{d\phi_t d\phi^*_t}{2 \pi i} \right] \frac{1}{\langle 
\phi_t | \phi_t \rangle} 
\langle \phi_t | \hat{T}^\dagger_t \,{\hat V}_t \hat{T}_{t+1}^{\vphantom{\dagger}} | \phi_{t+1} \rangle \label{zcn} 
\end{eqnarray}
where  a copy of the Fock space of the mesons has been introduced at each time slice. The  chemical potential
has disappeared because it is not active in a space of only mesons.
Explicitly
\begin{equation}
|\phi_t \rangle =
\exp\left(\sum_{\mathbf{x},K}\phi_K\left(t,\mathbf{x}\right){\vphantom{dagger}}
\hat{\Phi}^\dagger_{\mathbf{x}K}\left[U_t\right]\right)\,|0\rangle\, .
\end{equation}
{\em I remark that the structure functions $\Phi_{{\bf x},K}$ do not depend explicitly on time, but as they are functions
 of gauge fields, time will enter as a label of these fields.}
 
 In the evaluation of the trace on the fermionic Fock space
the only difference with respect to~\cite{Cara} is the presence of the operator ${\hat V}_t$. But I notice that
the product of operators ${\hat V}_t \, {\hat T}_{t+1} $ has an expression similar to that of
 ${\hat T}_{t+1}$
\begin{equation}
{\hat V}_t \, {\hat T}_{t+1}  =\exp [ -\hat{u}^{\dagger} \ln( e^{M_{t+1}} U_{0,t}^{\dagger})\, 
\hat{u} - \hat{v}^{\dagger}  
\ln( e^{M_{t+1}^T} U_{0,t}^T)  \hat{v} ]  \exp[\hat{v} N_t  \, \hat{u}] \,.
\end{equation}
Then evaluation of the trace over the Fock space proceeds exactly as in~\cite{Cara}
with the result 
\begin{equation}
\mathcal{Z}_{\mbox{mesons}} = \int [dU] \exp \left[S_G(U) \right]
\int\, \prod_t \left[ { d\phi_t d\phi_t^* \over 2 \pi i} \right] 
\exp\left[-S_{\mbox{mesons}} (\phi^*,\phi)\right]\, ,
\end{equation}
where
\begin{equation}
   \prod_t \left[ { d\phi_t d\phi_t^* \over 2 \pi i} \right] = \prod_{x,K} 
\left[ { d\phi_K(x) d\phi^*_{K}(x) \over 2\pi i } \right],
\end{equation}
and
\begin{equation}
S_{\mbox{mesons}} = \sum_t \mathrm{tr}_{-}\,\left[ \, - \ln \mbox{R}_t +  
\ln \mathcal{R}_t  +M_t^{\dagger}\right]
 \label{SM}\,.
\end{equation}
In the last equation
\begin{eqnarray}
\mbox{R}_t &= & \left(1 + {\mathcal F}^{\dagger} {\mathcal F}\right)_t^{-1}
\nonumber\\
\mathcal{R}_t  &=&  \left[ ( 1+{\mathcal F}^{\dagger}N )_{t+1} \,
 e^{M_{t+1}} U_{0,t}^{\dagger} \, e^{M_t^{\dagger}} 
  (1 + N^{\dagger} {\mathcal F} )_t \right.
  \nonumber\\
 &  &   \left.  
 + {\mathcal F} ^{\dagger}_{t+1} \, e^{-M_{t+1}}\,  U_{0,t}^{\dagger} \, e^{-M^*_t} 
{\mathcal F}_t  \,\right]^{-1} e^{ M_{t+1}} \, U_{0,t}^{\dagger}\,,
\end{eqnarray}
in which I set
\begin{equation}
{\mathcal F} = \phi^* \cdot  \Phi \,.
\end{equation}
Notice that the matrix ${\mathcal R}_t  $ involves gauge fields at time $t$ and $t+1$. 
The notation is somewhat different from that of~\cite{Cara}. 

It is remarkable that $S_{\mbox{mesons}}$ has been evaluated exactly~\cite{Cara}, so that 
{\it the only approximations in the partition function are the physical assumption of boson dominance  and the
form of the projector over the meson subspace.} Since the projector depends on the structure functions
$\Phi_{\bf{x},K}$, the effective action is a functional of these functions which are determined by
a variational calculation on the quantities of interest. In simple cases, like the four-fermion interaction
model, the variational calculation provides the exact form of the structure function. In QCD, unless
some analytic progress is made along a way similar to that of the four-fermion interaction model, one has
to adopt a trial expression.

In Ref.~\cite{Cara} an alternative, equivalent  form of the effective action was derived, which has a more
transparent interpretation. I will not report it here because I will also derive two forms of the effective
action at finite baryon density, but for the second one I will follow a somewhat different procedure.

\section{Quasiquarks and generalized Bogoliubov transformations}

In order to extend the formalism to QCD at finite baryon density, I must introduce in the  partition function 
states  with nonvanishing  baryon number. In the spirit of composites 
 dominance, I should then construct baryonic composites and define a projection operator on  the subspace of 
 mesons and baryons.
But in the present paper I will be satisfied by introducing, 
in addition to  mesons, quasiparticles states with quark quantum numbers, which I will
call quasiquarks (and quasiantiquarks). Such a space  obviously contains the
space of mesons and baryons. 

To avoid double counting, mesons and quasiparticles must satisfy a mutual 
 compositeness condition:  {\it Mesonic
states must be orthogonal to quasiquark-quasiantiquark states}. This constraint 
 has the physical meaning of the condition
$Z=0$ for bound states in the  Lehmann spectral representation of composite operators~\cite{Houa},
namely the   the condition required to introduce a bound state on the same footing as the constituents
in a Lagrangian.

 I will denote by 
${\hat \alpha}_i, \hat \beta_i$ 
 quasiparticles destruction operators. I will enforce
  the compositeness condition by requiring that quasiquark-quasiantiquark states
 annihilate  coherent states of mesons 
\begin{equation}
{\hat \alpha}_i |\phi \rangle =  {\hat \beta}_i |\phi \rangle = 0\,.
\end{equation}
This condition will prove crucial in the application to the 4-fermion model.

The compositeness condition in the above form can be solved exactly, even though composites coherent states 
do not satisfy exactly the basic property~(\ref{basic}) of coherent states of elementary bosons.
The  operators  ${\hat \alpha}_i, \hat \beta_i$ 
are obtained by a generalized Bogoliubov transformation~\cite{Bogo}
\begin{eqnarray}
{\hat \alpha}_i &=& \left[ \mbox{R}^{ 1 \over 2}\left( {\hat u} -  
 \,{\mathcal F}^{\dagger}  \, {\hat v}^{\dagger}\right) \right]_i 
\nonumber\\
{\hat \beta}_i &=& \left[  \left( {\hat v} +  
{\hat u}^{\dagger} \,{\mathcal F}^{\dagger} \right) (\mbox{R}\,')^{ 1 \over 2} \right]_i \, ,
\end{eqnarray}
where
\begin{equation}
\mbox{R}\,' = (1 + {\mathcal F} {\mathcal F}^{\dagger})^{-1} \,.
\end{equation}
 Bogoliubov introduced his tranformation  to construct a theory of superconductivity in the presence
of an electron-phonon interaction, and defines quasiparticles in terms of electron particle-holes states,
explicitly breaking the $U(1)$ symmetry related to fermion conservation. The above transformations
instead respect all symmetries, which is necessary with gauge interactions, thanks to the presence
of the bosonic field $\phi$.
The operators ${\hat \alpha}_i, \hat \beta_i $ and their Hermitean conjugates
\begin{eqnarray}
{\hat \alpha}_i^{\dagger} & = &\left[
\left( {\hat u}^{\dagger} - {\hat v} \, {\mathcal F}
\right)\mbox{ R}^{ 1\over 2} \right]_i
\nonumber\\
{\hat \beta}_i^{\dagger} & = &\left[(\mbox{R}\,')^{ 1\over 2} 
\left( {\hat v}^{\dagger} + {\mathcal F}\,
{\hat u}\right) \right]_i
\end{eqnarray}
 satisfy canonical commutation relations. As anticipated in the Introduction I include only mesons and
 quasiquarks in my variational
 space, but not quasiantiquarks. My variational space does not contain antibaryons, which are not
 expected to be important for not too high temperature and baryon density. In any
 case it will be clear that extension of the formalism to include quasiantiquarks should not
 present any significant difficulty.
 
  I define "coherent" states of quasiquarks and mesons
\begin{equation}
| \alpha, \phi \rangle = \exp (- \alpha \cdot {\hat \alpha}^{\dagger}) 
\exp ( \phi \,\cdot  {\hat \Phi}^{\dagger}) |0 \rangle
\end{equation}
where the $\alpha_i $ are Grassmann variables and 
\begin{equation}
\alpha \cdot {\hat \alpha} = \sum_i  \alpha_i  \, {\hat \alpha}_i  \,.
\end{equation}
 These states
 can be recast in the form
\begin{equation}
| \alpha, \phi \rangle =  \exp ({\hat u}^{\dagger} \, \mbox{R}^{-{ 1\over 2}} \alpha + 
 \phi \, \cdot {\hat  \Phi}^{\dagger}) |0 \rangle\
\end{equation}
which is more convenient for calculations.
The  operator which approximately projects on states of  mesons and quasiquarks is
\begin{equation}
\mathcal{P}_{m-q}= \int [d \alpha^* d \alpha]   \left[{ d \phi^* d \phi  \over 2 \pi i}\right]
\langle \alpha, \phi | \alpha, \phi \rangle^{-1}  \, | \alpha, \phi \rangle \langle \alpha, \phi | 
\end{equation}
where the measure is
\begin{equation}
\langle \alpha, \phi | \alpha, \phi \rangle^{-1} =\langle \alpha | \alpha \rangle^{-1}
\langle \phi |  \phi \rangle^{-1} =
\exp\{  \mbox{ tr}_{-}\ln \mbox{R}- \alpha^*  \cdot \alpha\} \label{measure}\,.
\end{equation}
If $\mathcal{P}_{m-q} $ is an approximate projector it  must  satisfy the equations
\begin{eqnarray}
&& \langle 0| \hat{\Phi}^{m_1} \hat{\alpha}^{n_1} \mathcal{P}_{m-q} \, 
(\hat{\alpha}^{\dagger})^{n_2}(\hat{\Phi}^{\dagger})^{m_2} | 0 \rangle  \simeq
\nonumber\\
&& 
\langle 0|\,  \hat{\Phi}^{m_1} \hat{\alpha}^{n_1}
 (\hat{\alpha}^{\dagger})^{n_2}(\hat{\Phi}^{\dagger})^{m_2} | 0 \rangle \varpropto 
\delta_{m_1,m_2}\delta_{n_1,n_2}\,. \label{ortho}
\end{eqnarray}
These equations are generated by the following ones
\begin{equation}
\langle  \phi_1 \alpha_1| \mathcal{P}_{m-q} | \phi_2\alpha_2 \rangle \simeq 
\langle  \phi_1 \alpha_1| \phi_2\alpha_2 \rangle \,, \label{ortho1}
\end{equation}
by taking derivatives with respect to the variables $\alpha_i, \phi_i$ and setting them equal to zero.
The left hand side of ~(\ref{ortho1}) is
\begin{eqnarray}
& & \langle \alpha_1, \phi_1|\mathcal{P}_{m-q}|\alpha_2, \phi_2 \rangle = 
\int [ d \alpha^* d \alpha] \left[{d\phi^* d\phi\over 2 \pi i}\right]
\exp \left \{ \mbox{tr}_{-}\left[ \, \ln \mbox{R} \right. \right.
\nonumber\\
&&+ \ln(1 + {\mathcal F}^{\dagger} {\mathcal F}_1 ) 
\left. \left. +  \ln(1 + {\mathcal F}_2^{\dagger} {\mathcal F}  )
\right] \right.
\nonumber\\
& &\left. + \alpha_1^*  \, \mbox{R}_1^{-{ 1\over2}}       \,
(1 + {\mathcal F}^{\dagger} {\mathcal F}_1 )^{-1} 
(1 + {\mathcal F}_2^{\dagger} {\mathcal F}  )^{-1}
\mbox{ R}_2^{-{ 1\over2}}   \, \alpha_2 \right\}\,.
\end{eqnarray}
This shows by inspection that the first member of~(\ref{ortho}) vanishes unless $m_1=m_2, n_1=n_2$.
 Evaluating the integral
in the above equation by the saddle point method, as done in~\cite{Cara} for $\mathcal{P}_{m}$, we see that 
$\mathcal{P}_{m-q}$ is approximately a projector if we assume
\begin{equation}
\mbox{tr}\left( \Phi^{\dagger} \Phi \right)^n \simeq \Omega^{-n+1}\,.
\end{equation}
 I remind that $\Omega$ is the index of nilpotency of $ \hat{\Phi}$.

\section{First form of the effective action at finite baryon density}

I follow the derivation of the effective action outlined in Section 2 for zero baryon density. 
I skip many intermediate steps because  calculations of this kind have been reported in any detail
in~\cite{Cara}, and can be easily repeated here by the help of the formulae collected in
Appendix~\ref{A}.
I start by evaluating  the matrix elements of the transfer matrix between coherent states
according to
\begin{eqnarray}
& &\langle \alpha_t, \phi_t |   \hat{T}^\dagger_t \, {\hat V}_t \exp(\mu  \, \hat{n}_B) 
\hat{T}_{t+1}
 | \alpha_{t+1}, \phi_{t+1} \rangle = \int [d\gamma^*d\gamma] [ d \delta^* d \delta]
 \nonumber\\
 & & \times  e^{  - \gamma^* \gamma  
 - \delta^* \delta  }
\langle \alpha_t, \phi_t |   \hat{T}^\dagger_t | \gamma \delta \rangle
 \langle \gamma \delta| {\hat V}_t \exp(\mu   \hat{n}_B) 
\hat{T}_{t+1}
 | \alpha_{t+1}, \phi_{t+1} \rangle \,.
\end{eqnarray}
The last factor is
\begin{eqnarray}
&& \langle \gamma \delta|\,  {\hat V}_t \exp(\mu   \hat{n}_B) 
\hat{T}_{t+1}
 | \alpha_{t+1}, \phi_{t+1} \rangle =
 {\det}_{-}( 1 + {\mathcal F}^{\dagger} N)_{t+1}
 \nonumber\\
&& \times \exp \left\{ \gamma^* U_{0,t} e^{-M_{t+1}}
( 1 + {\mathcal F} ^{\dagger} N )_{t+1}^{-1} 
 \left[ \, e^{\mu} \,{\mbox R}_{t+1}^{-{1\over 2}} \alpha_{t+1} +
{\mathcal F} ^{\dagger}_{t+1}
e^{-M_{t+1}} \,U_{0,t}^{\dagger} \delta^* \right]  \right\}.
\nonumber\\
\end{eqnarray}
A similar result for the other matrix element and integration over $ \gamma^*, \gamma , 
 \delta^*, \delta $ leads to the expression
\begin{eqnarray}
& &\langle \alpha_t, \phi_t |   \hat{T}^\dagger_t \, {\hat V}_t \exp(\mu  \, \hat{n}_B) 
\hat{T}_{t+1}
 | \alpha_{t+1}, \phi_{t+1} \rangle = {\det}_{-} \left(e^{- M_t^{\dagger}}\mathcal{R}_t^{-1} \right)
  \nonumber\\
 & &  \times 
 \exp\left(  \alpha_t^*  \, e^\mu   \,{\mbox R}_t^{-{ 1\over 2}} \,\,
  \mathcal{R}_t \,U_{0,t}\, e^{-M_{t+1}}
  {\mbox R}_{t+1} ^{ -{ 1\over 2}}  \, \alpha_{t+1}     \right) \,.
\end{eqnarray}

From the measure appearing in the definition of $ \mathcal{P}_{m-q}$, Eq.(\ref{measure}), I get the factor
\begin{equation}
\langle \alpha_t, \phi_t  | \alpha_{t}, \phi_{t} \rangle ^{-1}= 
  { \det}_{-}\,  {\mbox R}_t  \,  
 \exp\left( - \alpha_t^*  \cdot   \alpha_{t}     \right) \,.
\end{equation}
Putting these pieces together I get the effective action of mesons interacting with quasiquarks
\begin{eqnarray}
 S_{\mbox{mesons-quarks}} &= & S_{\mbox{mesons}} - \sum _t 
\alpha_t^*  \left[ - \alpha_t +\, e^\mu   \,{\mbox R}_t^{-{ 1\over 2}} \right. 
\nonumber\\
&& \left.  \,\times \,
  \mathcal{R}_t\,U_{0,t} \, e^{-M_{t+1}}
  {\mbox  R}_{t+1} ^{ -{ 1\over 2}}  \, \alpha_{t+1}  \right] \label{Smq}\,,
\end{eqnarray}
where $S_{\mbox{mesons}} $ is given by Eq.(\ref{SM}). I remind that that $\alpha$ is a 2-spinor with
the quark intrinsic quantum numbers.
It can be put in a more transparent form
\begin{equation}
 S_{\mbox{mesons-quarks}}= S_{\mbox{mesons}} -  s \sum _t  
   \alpha_t^*  ( \nabla_t -   \mathcal{H}_t )  \alpha_{t+1} 
   \end{equation}
by introducing the  lattice covariant derivative in the presence of a chemical potential and the lattice Hamiltonian
\begin{eqnarray}
\nabla_t \alpha_{t+1}&= & { 1\over s}  \left( e^{\mu} \, U_{0,t} \, \alpha_{t+1} - \alpha_t \right)
\nonumber\\
\mathcal{H}_t &=& { 1 \over s} \, e^{\mu}  \left[ U_{0,t} - {\mbox R}_t^{ -{ 1\over 2}}  \,\,
\mathcal{R}_t \, U_{0,t}
 e^{-M_{t+1}} \,{\mbox R}_{t+1}^{-{ 1\over 2}}  \right] \,.
\end{eqnarray}
The factor $s$ in the above equations takes the value 2 in the Kogut-Susskind regularization because
the quarks live on blocks, and 1 in the Wilson regularization. 
  I  notice that  the time derivative is not symmetric, so that this action does not give rise to
fermion doubling. Integrating over the Grassmann variables I get the purely bosonic effective action
\begin{equation}
S_{ \mbox{effective}} = - {\mbox{ Tr}}_{-}  
\ln\left(  {\mathcal R}^{-1}\, {\mbox R} \,
e^M\,
- e^{\mu} \,U_0 \, T_0^{(+)} 
  \right)  \label{Seffec}
\end{equation}
where I adopted the following notations:
All  matrices (with exception of $T_{\mu}^{(\pm)}$) which do not have a time lable are diagonal 
 in time with matrix elements
\begin{eqnarray}
(U_0)_{{\bf x}_1,t_1, {\bf x}_2,t_2}& =&
 \delta_{t_1,t_2} \delta_{{\bf x}_1, {\bf x}_2} U_{0,t_1}({\bf x}_1)
 \nonumber\\
{ \mathcal R}_{i_1,t_1, i_2,t_2} &=& \delta_{t_1,t_2}({ \mathcal R} _{t_1})_{i_1,i_2} 
\end{eqnarray}
while the matrix elements  of space-time translation operators are
\begin{equation}
\left( T^{(\pm)}_{\mu}  \right)_{x_1, x_2}= \delta_{x_2, x_1 \pm s \hat{\mu}} \,. 
\end{equation}
 $"{\mbox{ Tr}}" $ is the trace on all entries including  time, while I remind that  $"{\mbox{ tr}}" $
is the trace on intrinsic and spatial quantum numbers only.

\section{Application to a four-fermion interaction  model with Kogut-Susskind fermions}

To get insight in the above result and also to test it I apply it to the four-fermion interaction model
adopted as a test at zero fermion density~\cite{Cara}. It is a model  
in 3+1 dimensions regularized on a lattice with Kogut-Susskind fermions in the flavor basis (I do not 
know any formulation of the transfer matrix in the spin-diagonal basis
which  can be used in the present formalism).
 For each of the four Kogut-Susskind tastes there are $N_f$
degenerated flavors. Hence, the continuum limit will describe a theory
with $4N_f$ flavors. 
In the flavor basis the action reads  
\begin{equation}
{\mathcal S} =  \sum_x{}^\prime \sum_y{}^\prime \bar\psi(x) \,\left[ m \, 1\!\!1  \otimes 1\!\!1  \,
 + Q \right]_{x,y} \psi(y) + 
{ 1 \over 2}\,  { g^2 \over 4 N_f} \sum_x{}^\prime (\bar\psi(x)\psi(x))^2  \, 
\end{equation}
where $m$ is the mass parameter, $g^2$ the coupling constant,  $\psi$ the fermion fields and 
 $Q$  the  hopping matrix:
\begin{equation}
Q= \sum_\mu\gamma_\mu\otimes 1\!\!1
\left[ P_\mu^{(-)}\nabla_\mu^{(+)} + P_\mu^{(+)}\nabla_\mu^{(-)}\right]\, .
\end{equation}
The matrices to the left (right) of the symbol $\otimes$ act on Dirac (taste) indices. I denote by 
$\gamma$ and $t$ the matrices acting on 
these indices, respectively. The operators
\begin{equation}
P_\mu^{(\pm)}=\frac{1}{2}
\left[1\!\!1 \otimes 1\!\!1 \pm\gamma_\mu\gamma_5\otimes t_5t_\mu\right]\, .
\label{spin_projectors}
\end{equation}
are orthogonal projectors. The fermion fields are defined on blocks (see Appendix~\ref{B} for details). 
The right
and left derivatives $\nabla_{\mu}^{(\pm)}$ are given by 
\begin{equation}
\nabla^{(\pm)}_{\mu} = \pm { 1 \over 2} \left( T^{(\pm)}_{\mu} - 1  \right) \,.
\end{equation}
The factor $ 1/2$ is due to the fact that the operators $T_{\mu}$ translate by one block.
The model has a discreet chiral symmetry at $m=0$:
\begin{equation}
\psi\,\rightarrow\, -\gamma_5\otimes t_5 \,\psi\, , \hspace{1truecm}
\bar\psi\,\rightarrow\, \bar\psi\,\gamma_5\otimes t_5\, .
\end{equation}
To have an action bilinear in the fermion fields  a 
scalar field $\sigma(x)$ is introduced, whose integration generates the 
four-fermion coupling:
\begin{equation}
\mathcal{S}'=    \sum_x{}^\prime \sum_y{}^\prime \bar\psi(x) \,[  (m+ \sigma) 1\!\!1 \otimes 1\!\!1+ Q ]_{xy}\psi(y) 
+ \frac{4 N_f}{2g^2}\sum_x{}^\prime \,\sigma^2(x)  \, .
\end{equation}
The partition function now reads
\begin{equation}
\mathcal{Z}=\int[d\sigma][ d \overline{\psi} d \psi ] \,
 \exp\left[-  \mathcal{S}' \right] \, .
\end{equation}
Its restriction to fermion composites plus a fermion gas, without antifermions is 
\begin{equation}
\mathcal{Z}_{C-F}=\int[d\sigma] [ d \phi^* d \phi ] [ d \alpha^* d \alpha]\,
 \exp \left[ -  \frac{4 N_f}{2g^2}\sum_x{}^\prime \,\sigma^2(x)  -
 S_{C-F} \right] \, .
\end{equation}
$ S_{C-F} $ is given by Eq.(\ref{Smq}), in which one has to insert the expressions of the matrices
$M,N$ appropriate to  Kogut-Susskind  fermions in the flavor basis~\cite{Palu1}: The matrix $M$ 
is equal to zero and the matrix $N$ is reported in Appendix~\ref{B}.
Integration over the fermion fields gives the effective action
\begin{equation}
S_{ \mbox{effective}} = - {\mbox{ Tr}}_{-}\ln\left[ {\mathcal R}^{-1}\, 
\mbox{R} - e^{\mu} \, T_0^{(+)}   \, \right]  \,.
\end{equation}
Now I look for constant values of the fields  $\phi^*, \phi$ and $\sigma$ which make the action stationary.
 I put a bar over  constant fields and their functions. Then
 \begin{equation}
\overline{S}_{ \mbox{effective}} = - {\mbox{ Tr}}_{-}\ln\left[  \overline{\mathcal{R}}^{-1} \, 
 \overline{\mbox{R}} - e^{\mu} 
\, T_0^{(+)}  
 \right]  \,
\end{equation}
and  I can perform the sum over time getting
\begin{equation}
\overline{S}_{ \mbox{effective}}=-{ 1\over 2} L_0  \, \mbox{tr}_{-}  \left\{ \mu \, \theta \left[
 e^{\mu} -  \overline{\mathcal{R}}^{-1} \overline{{\mbox R}} \right] + 
\ln( \,\overline{{\mbox R}} \,  \overline{\mathcal{R}}^{-1}) \,
\theta \left[\overline{\mathcal{R}}^{-1} \, \overline{R} -  e^{\mu}  \, \,  
\right] \right\} \label{Sbar}\, ,
\end{equation}
where $\theta$ is the step function. For $\mu=0$ I recover the effective action  
derived~\cite{Cara}  at
 zero fermion density.

To determine the magnitude of the condensate 
 I must perform a variation  with respect to the boson fields $\overline{\phi}^*, \overline{\phi}$,
 and to determine the form factor of the composite a variation with respect to 
 the matrices ${\overline{\mathcal F}}^{\dagger}, {\overline{\mathcal F}}$. 
But $\overline{S}_{\mbox{effective}} $ does not depend on these
variables separately, it is a function of  ${\overline{\mathcal F}}^{\dagger}$ and 
${\overline{\mathcal F}} $. 
The saddle point equations with respect to these matrices 
for  $ e^{\mu}  < \overline{\mathcal{R}}^{-1} \,  {\overline{\mbox R}} $,
are identical to the ones for zero chemical potential. 
Using the result of~\cite{Cara}   I  then get
\begin{equation}
{\overline{\mathcal F}}^{\dagger} =  {N \over 2 H}
 \left( \sqrt{1 + H^2} +1 \right), \,\,\,
 e^{\mu}  < \overline{\mathcal R}^{-1} \,  {\overline {\mbox R}} \label{ff}\,,
\end{equation}
where
 \begin{equation}
 H= { 1\over 2} \sqrt{ N^\dagger N } = \sqrt{( m + \overline{\sigma})^2 - \bigtriangleup}
 \end{equation}
 with $ \bigtriangleup$ given by Eq.(\ref{laplascian}).
 I notice that $H$ differs from the  lattice Hamiltonian defined above
 \begin{equation}
 \overline{\mathcal H} = { 1\over 2} e^{\mu} \left(1-  
 {\overline {\mbox R}}^{-1} \, \overline{\mathcal R}\right)
 =e^{\mu} H \left( \sqrt{1 + H^2} - H \right),
 \end{equation}
 but they are equal in the formal limit of vanishing lattice spacing. In this limit I can
  rewrite  $S_{C-F} $ in the form
 \begin{equation}
S_{C-F} = S_C -2 \sum _t  
  \alpha_t^* \left[ \nabla_t^{(+)} -  (H -\mu)  \right] 
  \theta ( 2H - \mu)\alpha_t  \,. 
\end{equation}
For $m=\mu=0$, I recover the well known result that the fermionic
 system under consideration in the limit of $N_f \rightarrow \infty$ contains free fermions of mass 
 $\overline{\sigma}$ in addition to free bosons
 of mass $2 \, \overline{\sigma}$. I emphasize that the result recovered in this way is only formal.
 Indeed after adding one fermion  I should determine the new minimum of the action, namely
 the variation of the structure functions. But it can be justified in a concrete way by evaluating the difference 
 of $\overline{S}_{\mbox{effective}}$ given by Eq.{\ref{Eff}} below 
 at fermion numbers differing by one unit.
 
  Now I impose  the condition on the  fermion number which determines the chemical potential. From 
  Eq.(\ref{Sbar}) I get
 \begin{equation}
- { 2 \over L_0} {\partial  \over \partial \mu}\overline{S}_{\mbox{effective}}= 
\mbox{tr}_{-} \, \theta \left[ \exp\mu - \overline{\mathcal{R}}^{-1} \,{\overline{\mbox R}} 
\right] =\mbox{tr}_{-} \, \theta \left[ \exp\mu - 1 -2 \overline{\mathcal{H}} \right] = 
 n_F\,.
 \label{fermiondensity}
\end{equation}
 For $\mu<2 \, \overline{\sigma}$, $n_F=0$. {\it For $\mu > 2 \, \overline{\sigma}$,  quasifermions 
 occupy the  states from zero energy up to a maximum energy
 $E_{n_F}$ depending on the fermion number $n_F$}. 
 
  The effective action at the minimum takes the form
\begin{equation}
\overline{S}_{\mbox{effective}}= -L_0 \, \mbox{tr}_{-}\left\{  \ln \left(\sqrt{1+H^2}+
H \right)^2 
\theta \left(  2 \, \overline{\mathcal {H}} + 1- \exp\mu \right) \right\}\,. \label{Eff}
\end{equation}
Stationarity with respect to $\overline{\sigma}$ yields the gap equation which determines 
the masses and therefore the breaking of chiral invariance
\begin{equation}
 { 4 L_0 N_f \over g^2} \, \overline{\sigma}  =  - {\partial  \over \partial \overline{\sigma}}
 \overline{S}_{\mbox{effective}}=
    2 L_0 \, \overline{\sigma} \, \mbox{tr}_{-} \left\{ {1 \over H \sqrt{1+H^2}} \,\,
 \theta \left(  2 \, \overline{\mathcal {H}} + 1- \exp\mu\right) \right\}\label{gap}\,.
 \end{equation}
 Increasing the fermion density, namely the chemical potential,  quasifermions occupy 
 higher and higher energy states depleting the condensate,  until only the solution 
 $\overline{\sigma}=0$ remains and chiral invariance is restored.

\section{Second form of the effective action}

The expression~(\ref{ff}) of the form factors  is somewhat surprising, because they are increasing
functions of momentum. In~\cite{Cara}  a more natural form was deduced by performing a unitary transformation
in the fermionic Fock space and deriving the corresponding effective action. This transformation
changes the empty lattice into the fully occupied one and particles into holes. In this new
Fock space  the structure functions are decreasing functions of momentum, and in 
a polar representation their polar factor is equal to that of the Cooper pairs of the BCS
model of superconductivity.  

But in addition  the second form of the action
provided a test of consistency of the approximation for the projection operator $\mathcal{P}_m$. I
could follow the same path at nonzero baryon density, but instead I will get  
a similar result
in a different way. First I rearrange the trace in Fock space in the following way
\begin{equation}
\mathcal{Z} = \mathrm{Tr}^F\,\left\{{\hat V}_0 \,\exp(\mu \,n_B)  
\hat{T}_{1}^{\vphantom{\dagger}}\,
 \hat{T}^\dagger_1 \, {\hat V}_1 \,\,\exp(\mu \,n_B)  \cdots 
{\hat V}_{L_o-1}\exp(\mu \,n_B) \hat{T}_{0}^
{\vphantom{\dagger}}  \hat{T}^\dagger_0\,\right\} \,.
\end{equation}
Then I insert the projection operator $\mathcal{P}_{\mbox{m-q}}$ in the trace according to
\begin{eqnarray}
\mathcal{Z}'_{\mbox{mesons-quarks}} &=& \int [dU] \exp \left[S_G(U) \right]\mbox{Tr}^F  \left\{
\prod_{t=0}^{L_0-1} \right.
\nonumber\\
& & \left.\phantom{ \prod_{t=0}^{L_0-1}} \times
 \left(\mathcal{P}_{\mbox{m-q}} \exp(\mu \, \hat{n}_B) {\hat V}_t \, \hat{T}_{t+1} 
  \, \hat{T}^\dagger_{t+1} \right)  \right\} \,.
\end{eqnarray}
I emphasize that $\mathcal{Z}', \mathcal{Z}$ need not coincide with each other 
because $\mathcal{P}_{\mbox{m-q}}$
is not an exact projection operator, but the results obtained by the two forms should agree
within the approximation for $\mathcal{P}_{\mbox{m-q}} $. A comparison between these  
results  provides a check of its accuracy.

In the same way as for the first form of the effective action I evaluate the matrix elements 
\begin{eqnarray}
& &\langle  \alpha_t, \phi_t   | \exp(\mu \, \hat{n}_B){\hat V}_t \,
 \hat{T}_{t+1} \, \hat{T}^\dagger_{t+1} |     \alpha_{t+1}, \phi_{t+1}\rangle =
 \exp \left\{ - \mbox{tr}_{+} \mathcal{R}_t^{'} \right.
 \nonumber\\
 & &  \left. +  \,   \alpha_t^* \, e^{\mu} \, {\mbox R}_t^{-{ 1\over 2}} \, U_{0,t}
\,  e^{-M_{t+1}} \, \mathcal{R}_{t+1}^{'}   \,
 {\mbox R}_{t+1}^{-{ 1\over 2}} \, \alpha_{t+1} \right\}\,,
\end{eqnarray}
where
\begin{eqnarray}
\mathcal{R}_t^{'} &=&\left[1 +  \left( N_t + e^{-M_t}\,U_{0,t-1}^{\dagger} \, {\mathcal F}_{t-1} \,
U_{0,t-1}\, e^{- M_t} \right)^{\dagger} \right.
\nonumber\\    
& & \left.\times\left( N_t + e^{-M_{t}}\,
 {\mathcal F}_{t}  \,  e^{- M_{t}}\right) \right]^{-1}  e^{-M_t^{\dagger}} \,. \label{R}
  \end{eqnarray}
 Including the contribution~(\ref{measure}) from the measure I  get 
\begin{equation}
S^{'}_{\mbox{mesons-quarks}} = S^{'}{\mbox{mesons}}
- s \sum_t   \alpha^*_t \left( \nabla_t -
\mathcal{H}^{'}_t  \, \right) \alpha_{t+1}
\end{equation}
where
\begin{eqnarray}
S^{'}_{\mbox{mesons}} &=&\sum_t  \mbox{tr}_{-}\,  [ \,- \ln {\mbox R}_t +
\ln {\mathcal R}_t^{'} + M_t^{\dagger}] 
\nonumber\\
\mathcal{H}' &=& { 1 \over s} e^{\mu} \left[ U_{0,t} - {\mbox R}_t^{-{ 1\over2}} U_{0,t}
e^{-M_{t+1}} \, {\mathcal R}_{t+1}^{'} \,
\, {\mbox R}_{t+1}^{-{1 \over 2}} \right] \,.
 \nonumber\\
\end{eqnarray}
Integrating over $\alpha^*, \alpha$ I get the purely bosonic action
\begin{equation}
S_{\mbox{effective}}' =  - \mbox{Tr}_{-}  
\ln  \left[ - {\mbox R}  \, \, 
({\mathcal R}^{'})^{-1} e^{M}
 + e^{\mu} \,U_0 \,T_0^{(+)}  \, \right]\,.
\end{equation}
By exploiting the cyclic property of the trace it can be rewritten
\begin{equation}
S_{\mbox{effective}}' =  - \mbox{Tr}_{-} 
\ln  \left[  ({\mathcal R}^{'})^{-1} {\mbox R}  \, \, 
 e^{M}
 - e^{\mu} \,U_0 \,T_0^{(+)}  \, \right]\,.
\end{equation}
This expression differs from  $S_{\mbox{effective}} $, Eq.(\ref{Seffec}), by the replacement of
${\mathcal R}$ by ${\mathcal R}^{'}$.

I use this second form of the effective action for the four-fermion interaction model with Kogut-Susskind
fermions in the saddle point
approximation.
By means of the results of ref.\cite{Cara} I find 
\begin{equation}
{\overline {\mathcal F}}^{\dagger}= { N \over 2H} ( \sqrt{1+H^2}- H), \,\,\,
2 \, \overline{\mathcal{H}}' >  e^{\mu} -1\,.
\end{equation}
Now the structure function is a decreasing function of the constituent fermions energy. Using the above expression
 I find that 
\begin{equation}
  \overline{\mathcal{H}}' = \overline{\mathcal{H}}\,,
\end{equation}
so that the  results concerning  mass of the uncorrelated fermions and restoration of chiral
symmetry derived by the first form of the action are recovered.

I conclude this Section by an observation about the way the arguments
of Ref.\cite{Aloi} concerning the stability of numerical simulations might apply to the present 
effective action, comparing QCD with the four-fermion model.
In this model after linearization there is a field, the sigma field, which 
appears in the  matrix $N$ in the 
same position as the fermion mass, so that its expectation value provides in a  natural way a mass
to the fermion in the broken phase. In QCD there is no such field, but the chiral sigma field can play the
same role. Indeed  its expectation value appears in Eq.(\ref{R}) as an addendum to the 
matrix $N$ and therefore as an addendum to the bare quark mass. To the extent that high quark 
masses can stabilize numerical simulations, use of the present effective action should make
these calculations easier.

\section{Summary and outlook}

I  extended the formalism of composite boson dominance to 
the case of nonvanishing fermion number. This required a definition of fermion  and antifermion states in the 
presence of bosonic composites satisfying a compositeness condition to avoid double
counting. These fermion (antifermion) states are called quasifermions (quasiantifermions). 
Their definition is achieved by a 
generalized Bogoliubov transformation.

In the application to QCD I restricted myself to a space of mesons and quasiquarks, excluding
quasiantiquarks, which contains
the space of mesons and baryons.
Neglecting quasiantiquarks amounts to neglect virtual baryons-antibaryons, which is justified for
not too high temperature and baryon density. Obviously if one wants to investigate any
high temperature and/or baryon density phase transition quasiantiquarks must be included,
but I cannot foresee any obstruction in this extension.

I derived two forms of the effective action and 
I applied both of them to a four-fermi interaction model at zero temperature but
finite fermion density. I recovered in both ways the known results,
which provides a crosscheck of the approximation of the projection operator introduced to
restrict the fermionic Fock space in the partition function.  The discrete chiral invariance of the 
model (at zero fermion mass) is broken by composite boson
condensation and the spectrum of the broken phase contains, in addition to a composite  boson, 
a free fermion whose mass is half that
of the boson. Increasing the fermion density, quasifermions occupy the lowest energy
states up to an energy which increases with increasing density depleting the condensate,
until chiral symmetry is restored. The compositeness condition is crucial to get these results.
Since the action of the composite boson is known~\cite{Cara}, one could study the system also
at finite temperature and density. 

Possible applications of the present formalism include numerical studies of the evolution
 of the state of baryon matter with temperature and density and the associated phase transitions.
 In this connection I remind the way the arguments
of Ref.\cite{Aloi} concerning the stability of numerical simulations might apply:
the expectation value of the chiral sigma field appears in Eq.(\ref{R}) as an addendum to the 
matrix $N$ and therefore as an addendum to the bare quark mass. To the extent that high quark 
masses can stabilize numerical simulations, use of the present effective action should make
these calculations easier.
 
 Also exotic states of baryon matter can be explored. For instance 
  it is not difficult, as it will be shown in a separate paper,
  to introduce in the present formalism diquark states.
 Among   abnormal states of hadronic matter I would like to mention  the layered 
 spin-isospin phase \cite{Calo}. This is a state  with one-dimensional crystallization (which
 distinguishes it from usual pion condensation) in which layers of
 spin-up protons  and  spin-down neutrons  alternate with layers of spin-up neutrons and 
 spin-down protons. An investigation of a dynamical realization of such a phase in light deformed nuclei
 showed that the spin-isospin nucleon-nucleon interaction  is not sufficiently strong  to produce it~\cite{Rang},
while the critical density for a static phase in neutron stars has  been estimated~\cite{Bena}  to be
$3 \sim 4$ times normal nuclear density. A first principles calculation might nevertheless be worth while
to make an assessement of some simplifications done in the quoted works. 

To perform numerical simulations it 
is necessary to adopt a trial expression of the mesons structure functions, which should be a
function of gauge fields depending on
 temperature and baryon density, as suggested by the example of the four-fermion model.
To this end any analytical investigation of the effective action, for instance according
to an expansion in inverse powers of the index of nilpotency, might be of great help. In its absence,
the form of trial structure functions can be suggested by existing results about the spatial structure of hadrons,
of which a few examples can be found in~\cite{Velik}.

\vspace{0.5truecm}
\noindent 
\textbf{Acknowledgments}
I am grateful to  S. Caracciolo and G. DiCarlo for many fruitful discussions.
This work has been partially supported by EEC under the contract "Forces Universe" MRTN-CT-2004-
005104
\vspace{0.5truecm}

\appendix

\section{ Grassmann integrals and coherent states\label{A}}

If $|\alpha\rangle  $ is a fermionic coherent state
\begin{equation}
|\alpha\rangle = \exp (-\alpha \, \hat{u}^\dagger) |0\rangle
\end{equation}
then
\begin{equation}
\langle\alpha|\alpha\rangle = \exp ( \alpha^* \alpha) 
\end{equation}
and the identity can be written
\begin{equation}
\int [d\alpha d\alpha^*]\, \langle\alpha|\alpha\rangle^{-1} 
 |\alpha\rangle\langle\alpha| = 1 \,.
\end{equation}
I remind the fundamental property of coherent states
\begin{equation}
\hat{u} |\alpha\rangle =  \alpha  |\alpha\rangle 
\end{equation}
which implies the relations
\begin{eqnarray}
\langle\alpha\beta| \exp (\hat{v}N\hat{u}) |\gamma\delta\rangle & = & 
\exp (\delta N \gamma) \langle\alpha\beta|\gamma\delta\rangle = \exp (\delta N \gamma+\alpha^*\gamma +
\beta^*\delta)\\
\langle\gamma\delta| \exp( \hat{u}^\dagger {\mathcal F}^\dagger \hat{v}^\dagger)|0\rangle & =
 & \langle0| \exp( \hat{v}{\mathcal F}\hat{u}) |\gamma\delta\rangle^* = \exp( \gamma^*{\mathcal F}^\dagger \delta^*) \,.
\end{eqnarray}
With the help of these formulae one can compute matrix elements of the type
\begin{eqnarray}
\lefteqn{\langle \alpha \beta |e^{\hat{v} N \hat{u}} e^{\hat{u}^\dagger {\mathcal F}^\dagger \hat{v}^\dagger} |0\rangle}\\
&& = \int \left[\frac{d\gamma^* d\gamma  d\delta^* d\delta}{\langle\gamma\delta|\gamma\delta\rangle}\right] \langle \alpha \beta |e^{\hat{v} N \hat{u}}|\gamma\delta\rangle\langle\gamma\delta| e^{\hat{u}^\dagger {\mathcal F}^\dagger \hat{v}^\dagger} |0\rangle \\
&& = \int [d\gamma^* d\gamma d\delta^* d\delta] e^{-\gamma^*\gamma - \delta^*\delta +\delta N \gamma +\alpha^*\gamma + \beta^*\delta+\gamma^*{\mathcal F}^\dagger\delta^*}\\
&& =  \int [d\delta^* d\delta] \, e^{ - \delta^*(1+{\mathcal F}^* N^T)\delta  + \beta^*\delta - \delta^*{\mathcal F}^*\alpha^*}\\
&& =  \exp \left\{\mbox{tr}_{-} \ln(1+{\mathcal F}^* N^T)
-\beta^* (1+{\mathcal F}^* N^T)^{-1} {\mathcal F}^* \alpha^* \right\} \,,
\end{eqnarray}
by use of the identity
\begin{equation}
\int [d\alpha^* d\alpha]\,\exp( -\alpha^* A \alpha + J^*\alpha + \alpha^* J ) =\det A \,
\exp ( J^* A^{-1} J) \,.
\end{equation}

\section{The matrices $M,N$ of the transfer matrix\label{B}}

In this Appendix I report the expressions of the matrices $M, N$ appearing in the definition of the transfer matrix for the
Kogut-Susskind and  the Wilson regularization. Their   common feature is that they  depend only on the spatial link
variables.

\subsection{ Kogut-Susskind's regularization }

Kogut-Susskind fermions in the flavor basis are defined on hypercubes whose sides are  twice the basic lattice spacing. 
While in the text intrinsic quantum numbers and spatial coordinates were comprehensively  represented by one
index $i$, here I distinguish the spinorial index $\alpha=\{1,\ldots,4\}$, the taste index 
$a=\{1,\ldots,4\}$ and the flavour index i=$\{1,...,N_f\}$, while $x=\{t,x_1,\ldots,x_3\}$ is a 4-vector of {\em even} 
integer coordinates ranging in the intervals $[0, L_t-1]$ for the time component and $[0, L_s-1]$ for each of the spatial components.
I distinguish summations over  basic lattice and hypercubes according to
\begin{equation}
\sum_x{}^\prime := 2^d \sum_x \,.
\end{equation}
The projection operators over fermions-antifermion states are
\begin{equation}
P^{(\pm)}_0 = { 1\over 2} ( 1\!\!1 \otimes 1\!\!1 \pm \gamma_0 \gamma_5 \otimes t_5 t_0 ) \,.
\end{equation}
The relation between the variables $u,v$ and the quark $q$ field is
\begin{equation}
P^{(+)}_0 q ={ 1 \over 4} u, \,\,\, P^{(-)}_0 q = { 1 \over 4} v^\dagger \,.
\end{equation}
In the presence of the scalar field $\sigma$ and of gauge fields, neglecting an irrelevant constant,  $M=0$, 
while $N$ is~\cite{Palu1}
\begin{equation}
N =  -2  \left\{  ( m +\sigma)  \gamma_0  \otimes 1\!\!1  +  \sum_{j=1}^3  \gamma_0  \gamma_j   \otimes 1\!\!1  
 \left[  P^{(-)}_j   \nabla_j^{(+)}  + P^{(+)}_j \nabla_j^{(-)}
\right]  \vphantom{\sum_{j=1}^3} \right\}.\nonumber
\end{equation}
where
\begin{eqnarray}
 \nabla_j^{(+)} & = & { 1 \over 2}\left( U_j \,T^{(+)}_j  - 1 \right) \\
 \nabla_j^{(-)} & = &  { 1 \over 2}\left( 1- T^{(-)}_j  U_j^{\dagger}\right)
 \end{eqnarray}
 are the lattice covariant derivative as the $T_\mu^{(\pm)}$ are the forward and backward
translation operators of one block, that is of two lattice spacing in the original lattice, in the $\mu$  direction (with unit versor $\hat{\mu}$)
\begin{equation}
[T_\mu^{(\pm)}]_{x_1,x_2 }=\delta_{x_2,x_1 \pm 2\hat{\mu}}\, .
\end{equation}
I set 
\begin{equation}
 N^{\dagger} N = 4 H^2 \,. 
\end{equation}
 In the absence of gauge fields
\begin{equation}
H^2 =  ( m + \sigma) ^2 -  \Delta
\end{equation}
with
\begin{equation}
\Delta = { 1 \over 4}  \sum_{i=1,3} \left( T_i^{(+)}+ T_i^{(-)} -2 \right) \label{laplascian}
\end{equation}
The eigenvalues of $H^2  $ are therefore the fermion energies
\begin{equation}
E_q^2 =  m^2  + \tilde{p}^2 \, , \label{energy}
\end{equation}
where  momentum component $\tilde{p}^2_i $ is 
\begin{equation}
\tilde{p}^2_i = { 1\over 2} ( 1  - \cos 2\, p_i)\, . \label{momentum}
\end{equation}
and 
\begin{equation}
\tilde{p}^2 = \sum_{i=1}^3 \tilde{p}^2_i
\end{equation}

\subsection{ Wilson's regularization }

The projection operators over fermions-antifermions are
\begin{equation}
P^{(\pm)}_0 = { 1\over 2} ( 1 \pm \gamma_0 ) \,.
\end{equation}
in a basis in which $\gamma_0 = \mbox{diag}(1,1,-1,-1)$.

The relations between the quark field $q$ and its upper and lower components $u,v$  are
\begin{equation}
P^{(+)}_0 q = B^{- { 1 \over 2}}\, u, \,\,\, P^{(-)}_0 q = B^{- { 1 \over 2}} v^\dagger\,,
\end{equation}
where
\begin{equation}
B = 1 -  K \sum_{j=1}^3 \left (U_j T^{(+)}_j + T^{(-)}_j U^\dagger_j  \right) \gamma_j 
 \end{equation}
and $K$ is the hopping parameter.
The matrices $M, N$ are
\begin{eqnarray}
M&=& - { 1\over 2} \ln \left( {B \over 2K} \right)
\nonumber\\
N &= & 2 K \, B^{- { 1 \over 2}}\,c \, B^{- { 1 \over 2}} \,,
\end{eqnarray}
where
\begin{equation}
c = { 1\over 2}  \sum_{j=1}^3 i \left ( U_j  \, T^{(+)}_j - \, T^{(-)}_j U^\dagger_j  \right)\,
\sigma_j \,.
\end{equation}

\end{document}